# Highly efficient acousto-optic modulation using nonsuspended thin-film lithium niobate-chalcogenide hybrid waveguides

Lei Wan[1,#,*], Zhiqiang Yang[2,#], Wenfeng Zhou[1,#], Meixun Wen[1], Tianhua Feng[1], Siqing Zeng[2], Dong Liu[2], Huan Li[3], Jingshun Pan[2], Ning Zhu[4], Weiping Liu[1] & Zhaohui Li[2,5,*]

[1]*Department of Electronic Engineering, College of Information Science and Technology, Jinan University, Guangzhou 510632, China*
[2]*Guangdong Provincial Key Laboratory of Optoelectronic Information Processing Chips and Systems, Sun Yat-sen University, Guangzhou 510275, China*
[3]*State Key Laboratory for Modern Optical Instrumentation, College of Optical Science and Engineering, International Research Center for Advanced Photonics, Zhejiang University, Zijingang Campus, Hangzhou 310058, China*
[4]*Institute of Semiconductor Science and Technology, Guangdong Engineering Technology Research Center of Low Carbon and New Energy Materials, South China Normal University, Guangzhou 510631, China*
[5]*Southern Marine Science and Engineering Guangdong Laboratory (Zhuhai), Zhuhai 519000, China*
[#]*These authors contributed equally to this work.*
*\*wanlei@jnu.edu.cn*
*\*lzhh88@sysu.edu.cn*

**Abstract**

A highly efficient on-chip acousto-optic modulator, as a key component, occupies an exceptional position in microwave-to-optical conversion. Homogeneous thin-film lithium niobate is preferentially employed to build the suspended configuration forming the acoustic resonant cavity to improve the modulation efficiency of the device. However, the limited cavity length and complex fabrication recipe of the suspended prototype restrain further breakthrough in the ME and impose challenges for waveguide fabrication. In this work, based on a nonsuspended thin-film lithium niobate-chalcogenide glass hybrid Mach–Zehnder interferometer waveguide platform, we propose and demonstrate a built-in push-pull acousto-optic modulator with a half-wave-voltage-length product $V_\pi L$ as low as 0.03 V cm, presenting a modulation efficiency comparable to that of the state-of-the-art suspended counterpart. Based on the advantage of low power consumption, a microwave modulation link is demonstrated using our developed built-in push-pull acousto-optic modulator. The nontrivial acousto-optic modulation performance benefits from the superior photoelastic property of the chalcogenide membrane and the completely bidirectional participation of the antisymmetric Rayleigh surface acoustic wave mode excited by the impedance-matched interdigital transducer, overcoming the issue of amplitude differences of surface acoustic waves applied to the Mach–Zehnder interferometer two arms in traditional push-pull acousto-optic modulators.

## Introduction

The acousto-optic (AO) interaction is a multiphysics coupling process in which radiofrequency (RF)-driven acoustic waves in a medium can change the localized refractive index of the waveguide, thereby manipulating photons[1-8], which has stimulated ingenious applications in many areas, such as coherent quantum transduction[9], nonreciprocal light transmission[10, 11], modulation[12], frequency shifting[13], signal processing[14, 15], beam deflection[16, 17], and filtering[18]. Presently, the traditional AO devices based on bulk crystal materials have weak energy confinement for both photons and phonons, leading to a low AO interaction strength[19]. Compared with bulk materials, photonic integrated circuits (PICs) allow surface acoustic waves (SAWs) to be well confined within a thin film used to disturb the guided light waves, exhibiting a high energy overlap within the wavelength scale. Therefore, SAWs as an effective tool are desirable to achieve a very high AO modulation efficiency (ME) and can further enable diversity of functionalities in PICs.

The generation of SAWs relies on interdigital transducers (IDTs) placed over the thin-film (TF) piezoelectric materials in PICs. SAW-coupled high-performance AO modulation requires a low-loss optical waveguide and a high-efficiency IDT, which are simultaneously integrated into an on-chip optoelectronic platform by judiciously engineering the configurations of the optical and acoustic components and the relative position between them. With the development of fabrication technologies for TF piezoelectric materials, on-chip AO modulators have been accordingly demonstrated by homogeneously integrating waveguides and IDTs in the same optical membranes, such as gallium arsenide (GaAs)[20, 21], polycrystalline aluminum nitride (AlN)[5], and TF lithium niobate (TFLN)[22-24]. Particularly, as one of the most promising AO interaction platforms, TFLN provides great potential for realization of high-performance AO modulators due to its superior advantages in piezoelectric transduction and electro-optical conversion[25-30]. However, limited by the low optomechanical coupling coefficients, weak AO MEs have become one of the bottlenecks for microwave-to-optical conversion in 5G/6G and emerging quantum signal processing applications.

To meet the challenges arising from the AO ME, the configurations of acoustic cavities are commonly considered to enhance the amplitudes of acoustic waves. In nonsuspended TFLN, acoustic resonant cavities consisting of a pair of metal gratings are engineered to increase the SAW and AO interaction[23]. In fact, the low mechanical reflectivity and long grating dimension of the reflectors increase the losses of acoustic waves, leading to low acoustic Q-factors and large $V_\pi L$. To significantly improve the AO ME, a free-standing LN thin film as a feasible solution is preferred to form a standing wave between two open slits with a distance of tens of μm so that the excited weak acoustic wave can be enormously amplified to enhance the overlap factor between the optical and acoustic modes. However, the short distance along the propagation direction of the acoustic wave seriously limits the number of IDT fingers, resulting in suboptimal impedance matching and large microwave reflection[22, 24]. In addition, the construction of a suspended LN acoustic resonator and the position of the sensitive optical waveguide over the suspended acoustic resonator raise stringent requirements on the sophisticated fabrication recipes, especially for heterogeneous-integration waveguides. How to acquire a high AO ME without a suspended acoustic resonator has become an important topic for microwave-to-optical conversion.

Generally, the IDT of an AO modulator with a Mach–Zehnder interferometer (MZI) configuration is placed outside the two arms of a nonsuspended waveguide to modulate the refractive index of a single arm (SA). Since the SAW excited by the IDT propagates in two opposite directions, only 50% of the effective energy within the total converted acoustic wave can ideally reach the waveguide and participate in the interaction with light waves under the SA modulation mechanism, resulting in a quite low ME. Although push-pull MZI modulation configurations (odd multiple of half the acoustic wavelength) have been proposed by carefully designing the distance between the MZI two arms and optimizing the structures of metal grating reflectors, the low reflectivity of reflectors and the reconstructed acoustic mode introduced by the SAW reflection and interference originating from waveguide sidewalls inevitably lead to much more SAW energy being wasted and cannot enable an ME with a twofold enhancement, defeating the initial purpose of the double arm (DA) modulation[23, 31]. More specifically, a small error in the distance between the two arms may cause a very large difference in the DA modulation effect, which brings great challenges for the device fabrication tolerance.

To overcome the above problems, a novel built-in push-pull AO modulation configuration based on the antisymmetric SAW mode, for which the optimized IDT is placed inside the two arms of the designed MZI waveguide, is theoretically proposed and experimentally demonstrated based on nonsuspended TFLN without an acoustic cavity in our work. This configuration can suitably relax the fabrication tolerances of the optical waveguide and the IDT, enabling a two times enhanced ME compared with the SA configuration due to the employment of 100% of the bidirectional acoustic energy in principle. In addition, considering that amorphous chalcogenide glasses (ChGs) as soft infrared waveguide materials have a better photoelastic (PE) coefficient, nearly two times that of TFLN[32-35], the TFLN-ChG hybrid waveguide platform is adopted to facilitate improvement of the AO ME. Most of the optical energy is designed to be confined in the ChG rectangular waveguide to take full advantage of the dominant PE effect. Proper engineering of the geometry of the nonsuspended TFLN-ChG hybrid waveguide combined with a precisely defined IDT can not only significantly increase the DA ME in the push-pull MZI configuration but also avoid direct etching of TFLN, presenting potential prospects in microwave-to-optical conversion. In the experiments, the $V_\pi L$ of our AO modulator with the DA configuration based on the nonsuspended TFLN-ChG hybrid MZI is demonstrated to be as low as 0.03 V cm, which is one order of magnitude smaller than that of the nonsuspended homogeneous TFLN counterpart[24]. To the best of our knowledge, this is the first built-in push-pull modulator implemented in on-chip AO devices in which the ME exceeds that of AO modulators with acoustic cavities[22-24]. To further present the low power consumption feature, we characterize the optical and RF sidebands of the device and verify the efficient transmission of an on-off keying (OOK) modulated signal. Combined with the simple fabrication processes and high-performance ME, our built-in push-pull AO modulator is expected to show excellent characteristics in on-chip RF-driven optical isolators[36] and integrated analog optical computing systems[37].

## Results
### Device design

A schematic diagram of the proposed device is shown in Fig. 1a. The AO modulator made of an MZI waveguide and an IDT inside the MZI two arms is constructed on a nonsuspended TFLN-ChG hybrid waveguide platform. X-cut LN on insulator (LNOI) with a thickness of 400 nm is chosen to excite a bidirectional Rayleigh SAW using the piezoelectric effect and IDT. The MZI ridge hybrid waveguide is composed of a $Ge_{25}Sb_{10}S_{65}$ (one of the ChGs) rectangular waveguide and a TFLN slab because the refractive index of amorphous $Ge_{25}Sb_{10}S_{65}$ ($n$ = 2.23) is very close to the refractive index of the LN crystal ($n_e$ =2.13) at 1550 nm. The similar optical refractive indices of $Ge_{25}Sb_{10}S_{65}$ and LN enable the fundamental transverse-electric (TE) mode to be simultaneously confined in the $Ge_{25}Sb_{10}S_{65}$ and LN layers, as shown in Fig. 1b.

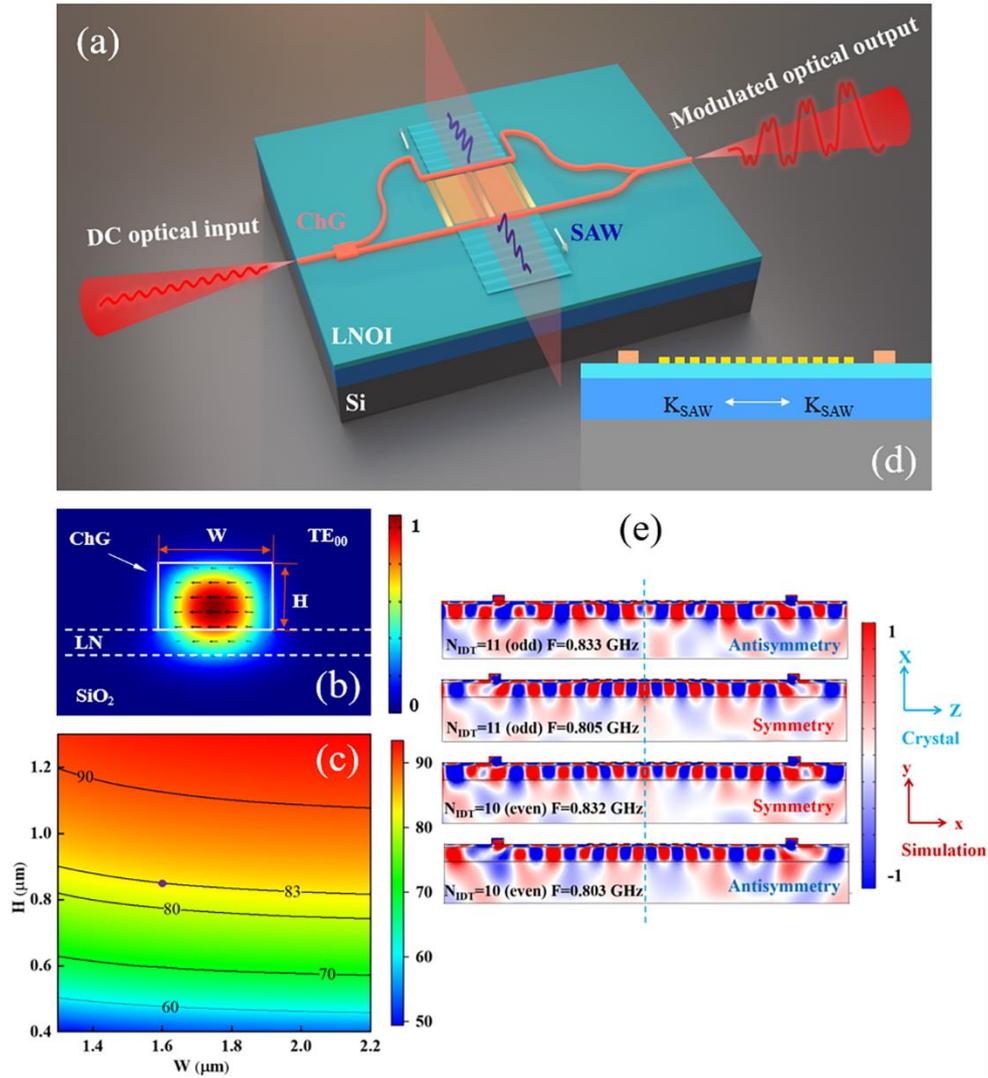

**Fig. 1 Design of a built-in push-pull AO modulator based on a nonsuspended TFLN-ChG hybrid waveguide platform. a** Schematic diagram of the proposed device. The MZI waveguide is etched on the top $Ge_{25}Sb_{10}S_{65}$ film (orange), which is deposited over the LNOI wafer (light blue). The IDT made of Au electrodes (yellow) is evaporated in the region between the two arms. In principle, the direct current (DC) optical input is modulated into a distorted sinusoidal time-domain signal via a varied SAW. **b** Electric field of the fundamental TE mode. W and H are the width and height of the ChG rectangular waveguide. **c** Relation between the energy confinement factor Γ in the ChG waveguide and waveguide geometry. **d** Cross-sectional view of the proposed device. The colors of the waveguide materials are the same as in **a**. **e** Numerical simulation results of the dominant $S_{xx}$ strain components of the SAW modes in the heterogeneous-integration waveguide platform with the DA modulation configuration. The upper two pictures are the antisymmetric and symmetric acoustic modes corresponding to 5.5 pairs of IDT fingers ($N_{IDT}$ = 11), and the lower two are the symmetric and antisymmetric acoustic modes corresponding to 5 pairs of IDT fingers ($N_{IDT}$ = 10). Color scale bars are normalized.

To determine the structural parameters of the built-in push-pull AO modulator, we conduct numerical simulations for the optical and acoustic modes in the nonsuspended TFLN-ChG hybrid waveguide platform. Herein, the thickness of TFLN is fixed at 400 nm. The geometries of the

ChG rectangular waveguides are varied to engineer proper parameters to adequately employ their advantage in the PE property. Figure 1c shows that the energy confinement factors of the fundamental TE modes in the ChG rectangular waveguides greatly depend on H rather than W. As H increases, the mode energy gradually concentrates in the ChG material. In our experiments, an H of 850 nm and a W of 2.05 μm are set to support the AO interaction because an overlarge sidewall height would bring increased scattering loss. In the AO interaction area, a taper is added to convert the waveguide width from 2.05 μm to 1.6 μm to match the half wavelength of the acoustic wave ($\Lambda$ = 3.2 μm), achieving maximum AO overlap. The $\Gamma$ of the 1.6 μm-wide ChG rectangular waveguide supporting the $TE_{00}$ mode is calculated to be 83%, which is highlighted by the purple dot in Fig. 1c. Optimized multimode interference (MMI) coupler and Y-combiner are placed at the input and output ports of the hybrid MZI to uniformly route the photon energy while decreasing the insertion loss (IL) of the device. Four ultracompact 90° ChG bending waveguides with an effective radius of 10 μm are inversely designed based on free-form curves[38]. The loss of each 90° bending waveguide is experimentally measured to be approximately 0.25 dB. The introduction of an ultracompact 90° bending waveguide can not only facilitate precise control of the distance between the MZI arms and flexibly designed IDT in the interaction area, avoiding the acoustic wave dissipation induced by a long propagation distance, but also decrease the size of the device while maintaining the proper IL.

Figure 1d shows a cross-sectional image of the TFLN-ChG hybrid MZI-based AO modulator in the interaction area, denoted by the light red plane in Fig. 1a. To reveal the acoustic mode excited by the built-in IDT in the two-arm waveguides, we simulate the dominant $S_{xx}$ strain components at an acoustic frequency (F) of nearly 0.833 GHz and 0.805 GHz, corresponding to 5.5 (finger number $N_{IDT}$ = 11, odd) and 5 ($N_{IDT}$ = 10, even) pairs of IDT fingers, respectively, as presented in Fig. 1e. Herein, in contrast to the experimental IDT, a reduced finger number is set in the simulation to reasonably reflect the acoustic mode distribution because the finger number difference does not influence the sign relation of the acoustic mode in the MZI two arms. The IDT with an odd number of fingers yields an antisymmetric strain field in the two-arm waveguides at 0.833 GHz, but at the low frequency of 0.805 GHz, the IDT with the same configuration exhibits a symmetric acoustic mode. In contrast, the IDT with an even number of fingers generates a symmetric strain field at 0.832 GHz, and the same IDT exhibits an antisymmetric acoustic mode at 0.803 GHz. This means that the signs of the strain fields in the two-arm waveguides strongly depend on the finger number of the IDT at the specific acoustic frequency. The simulation results of the $S_{ZZ}$ strain component have similar variations (see Supplementary note 1). By reasonably engineering the structure of the IDT with an odd number of fingers, opposite refractive index changes can be realized in the two-arm waveguides, satisfying the requirement of the push-pull AO modulator in principle. From the perspective of impedance matching, we design the IDT with 50.5 pairs of fingers ($N_{IDT}$ = 101) to efficiently excite the SAW in the following experiments.

**Device characterization**

Figure 2a shows optical microscopy images of the fabricated A device. For reference, two identical IDTs are placed inside and outside the two-arm waveguides, forming the DA and SA modulation configurations, respectively. To ensure that the SAW excited by the outside IDT only interacts with the lower waveguide, we pattern an unetched ChG patch along the propagation path of the SAW. Due to the strong absorption of ChG material for the SAW, the acoustic wave cannot reach the upper arm of the MZI. The three electrodes of the built-in IDT are ground-signal-ground. The widths of the ground and signal electrodes are set to 60 μm and 40 μm, respectively, and the aperture width is 2×60 μm. Figures 2b, c, and d are zoomed-in microscope images of the 1×2 MMI coupler, ultracompact 90° bending waveguide, and Y-combiner, respectively. The length and width of the MMI coupler are designed to be 97 and 11.5 μm, respectively. Figures 2e and 2f show scanning electron microscopy (SEM) images of the modulated waveguide in a cross-sectional view and IDT details. The width of the ChG waveguide is measured to be 1.6 μm, and the actual sidewall angle is estimated to be 89.6°. The gaps between the IDTs and two-arm waveguides are designed to be 11.8 μm. The width of each IDT finger is 0.8 μm.

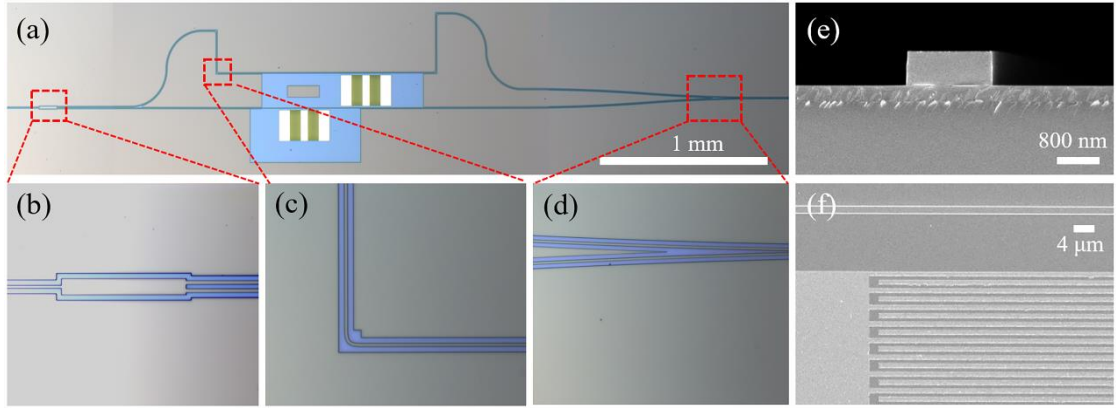

**Fig. 2 Configuration of the built-in push-pull AO modulator over the nonsuspended TFLN-ChG hybrid waveguide platform. a** Microscope image of a TFLN-ChG hybrid MZI-based AO modulator. The MZI is designed to allow careful switching between the SA and DA modulation mechanisms. **b** Zoomed-in microscope images of the 1×2 MMI coupler, **c** 90° bending waveguide, and **d** Y-combiner. **e** Cross-sectional SEM image of the modulated waveguide. **f** SEM image of the modulation area including the IDT and waveguide.

**Microwave-to-acoustic conversion**

Figure 3a shows the schematic diagram of the experimental setup of the on-chip AO modulator. We experimentally characterize the MZI-based AO modulator using a tunable C-band laser, a vector network analyzer, and a photoreceiver that features a sensitivity of 800 V/W. A fiber polarization controller is used to adjust the fundamental TE mode fed into the waveguide, and a pair of lens fibers with a mode field diameter of 3 μm are utilized for the edge coupling between the input and output light and device. The fiber-to-fiber IL of the AO modulator is measured to be 16 dB, and the edge-coupling loss is 5.5 dB/facet, illustrating an on-chip loss of 5 dB. The normalized optical transmission spectrum of the AO modulator with the SA modulation configuration (B device) presents periodic interference with a free spectral range of nearly 1.2 nm, as shown in Fig. 3b, which is consistent with the natural optical path difference of the designed MZI. As depicted in the transmission spectrum, the extinction ratio (ER) of the device is approximately 17 dB at approximately 1573 nm.

We evaluate the microwave-to-acoustic conversion by measuring the microwave reflection coefficient ($S_{11}$) spectrum of the fabricated IDT, which exhibits two resonance dips within the 0.25–1.25 GHz range, as shown in Fig. 3c. The deep reflection valley is located at 0.844 GHz, and $1-|S_{11}|^2$ is calculated to be 98.5%, which represents the ratio of the RF power loaded on the IDT to the input RF power. The experimentally observed dominant acoustic wave at 0.844 GHz belongs to the antisymmetric Rayleigh SAW, agreeing well with the acoustic mode displayed in the top panel of Fig. 1e. The slight variation in frequency may be caused by the fabrication deviation and finger number difference of the IDT in the two cases. The Smith chart shows that the designed IDT greatly satisfies the impedance matching condition at 0.844 GHz, as emphasized by the large circle in Fig. 3d. Accordingly, the complex impedance of the IDT at 0.844 GHz is measured to 56.99-$j$11.11 Ω, approaching the standard 50 Ω. To precisely estimate the microwave-to-acoustic conversion efficiency, a Butterworth-Van Dyke (BVD) model is introduced to reveal the energy consumption in load $Z_a$, which is the available acoustic energy in the DA modulation configuration, by fitting the nonresonant normalized complex impedances in the Smith chart[39], as denoted by the red dotted line in Fig. 3d. Herein, $R_s$ accounts for the contact resistance between the RF probe and IDT fingers, and $R_1$ and $C_e$ account for the effective leakage resistance and capacitance in the IDT region, respectively. The detailed calculation processes are described in Supplementary note 2. As a result, the microwave-to-acoustic conversion efficiency of the designed IDT is calculated to be 0.96 at 0.844 GHz (see the red line of Fig. 3c). Compared with the suspended TFLN proposed in Refs. 22 and 24, the flexible fabrication of the proposed IDT over the nonsuspended TFLN ensures highly efficient electromechanical conversion in our experiments.

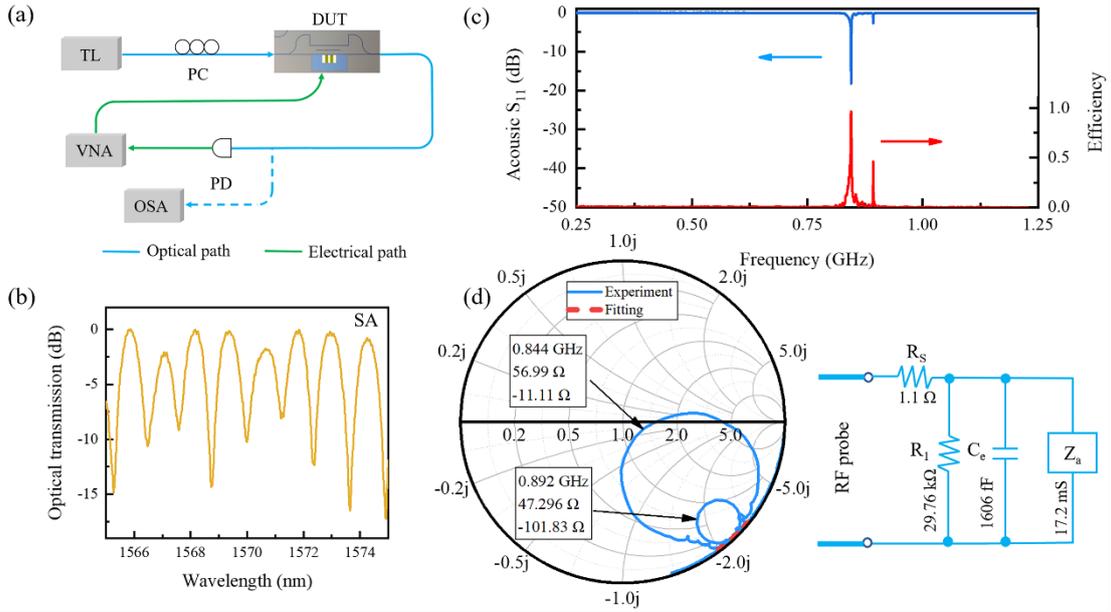

**Fig. 3 Characterization of microwave-to-acoustic conversion in the SA-modulated AO modulator. a** Schematic diagram of the device measurement system. **b** Optical transmission spectrum of an MZI-based AO modulator with the SA modulation configuration. **c** $S_{11}$ reflection spectrum and conversion efficiency of the fabricated IDT. **d** Smith chart of the fabricated IDT and corresponding effective circuit model. TL: tunable laser, VNA: vector network analyzer, DUT: device under test, PC: polarization controller, PD: photodiode, OSA: optical spectrum analyzer.

## Comparison of AO modulation

We characterize the AO modulation of the device by the opto-acoustic $S_{21}$ spectrum, where driving Port 1 of the VNA is connected to the IDT and detecting Port 2 is connected to the photoreceiver (see Fig. 3a). The $S_{21}$ spectrum shown in Fig. 4a has two significant peaks in the frequency range of 0–2 GHz, indicating that the microwave-to-optical conversion is enhanced at these frequencies. The strong opto-acoustic response is positioned at 0.844 GHz, corresponding to the sharp dip displayed in the $S_{11}$ spectrum (Fig. 3c). Compared with the B device with the SA modulation configuration, the built-in push-pull AO modulator (C device) obtains a nearly 12 dB improvement in $S_{21}$ under an RF power of 0 dBm, greatly presenting the effectiveness of DA modulation. The enhanced $S_{21}$ benefits from the increase in the ER at approximately 1561 nm in the transmission spectrum of the C device, as shown in Fig. 4b.

To further quantify the AO modulation characteristics, we extract the $V_\pi$ of the proposed nonsuspended MZI-based AO modulator from the experimental measurement of the $S_{21}$ spectrum by[17]

$$V_\pi = \frac{\pi R_{PD} I_{rec}}{|S_{21}|}, \qquad (1)$$

where $R_{PD}$ is the sensitivity of the photoreceiver and $I_{rec}$ is the DC optical power at the bias point with a $0.5\pi$ phase change. By choosing the bias point in the transmission spectrum, $I_{rec}$ is measured to be -23 dBm (-24 dBm) for the DA-modulated (C device) (SA-modulated, B device) AO modulator. From the $S_{21}$ spectrum, the DA-modulated (SA-modulated) AO modulator has an $S_{21}$ of -46 dBm (-58.8 dBm) at 0.844 GHz. The $V_\pi$ of our DA (SA) modulator is thus calculated to be 2.5 V (8.68 V), corresponding to $V_\pi L$ = 0.03 V cm (0.1 V cm) due to the modulation length of 120 μm, which shows that the built-in push-pull AO modulator has a threefold enhancement in the ME compared with the SA modulator. To the best of our knowledge, this is the state-of-the-art on-chip AO modulator based on TFLN with a superior ME. The excellent ME is attributed to the highly efficient employment of a bidirectionally propagating SAW with sufficient high energy combined with the outstanding PE property of the ChG material. Detailed comparisons of modulation characteristics are presented in Table 1. To strictly demonstrate the contribution of the proposed DA modulation configuration, SA- and DA-modulated configurations are simultaneously integrated with a TFLN-ChG hybrid MZI to alternately conduct measurements with the same optical transmission spectrum and bias point (A device), as shown in Fig. 2a. The results definitely confirm the relation of a two times enhancement in the ME (see Supplementary note 3), attributed to the ideally antisymmetric acoustic mode distribution in the two-arm waveguides excited by the built-in IDT.

To further reveal the modulation characteristics of the device, we change the experimental setup to observe optical sidebands, as shown in Fig. 3a. Herein, the output light is directly injected into the high-precision spectrometer, and the frequency of the RF signal is set to 0.844 GHz. When the RF power is 15 dBm, a weak third-order sideband appears in the spectrum (Fig. 4c) when using the push-pull AO modulator. The symmetric sideband spectrum closely depends on the choice of the bias point. More importantly, with the gradual increase of the RF power, we could expect to observe many more optical sidebands. As described in Ref. 37, the periodic sidebands can be regarded as frequency lattices, which may have potential in integrated analog optical computing. Meanwhile, the ER of the MZI is accordingly decreased (see Supplementary note 4), which may be caused by the enhanced acoustic wave modulation-induced energy dissipation in the MZI.

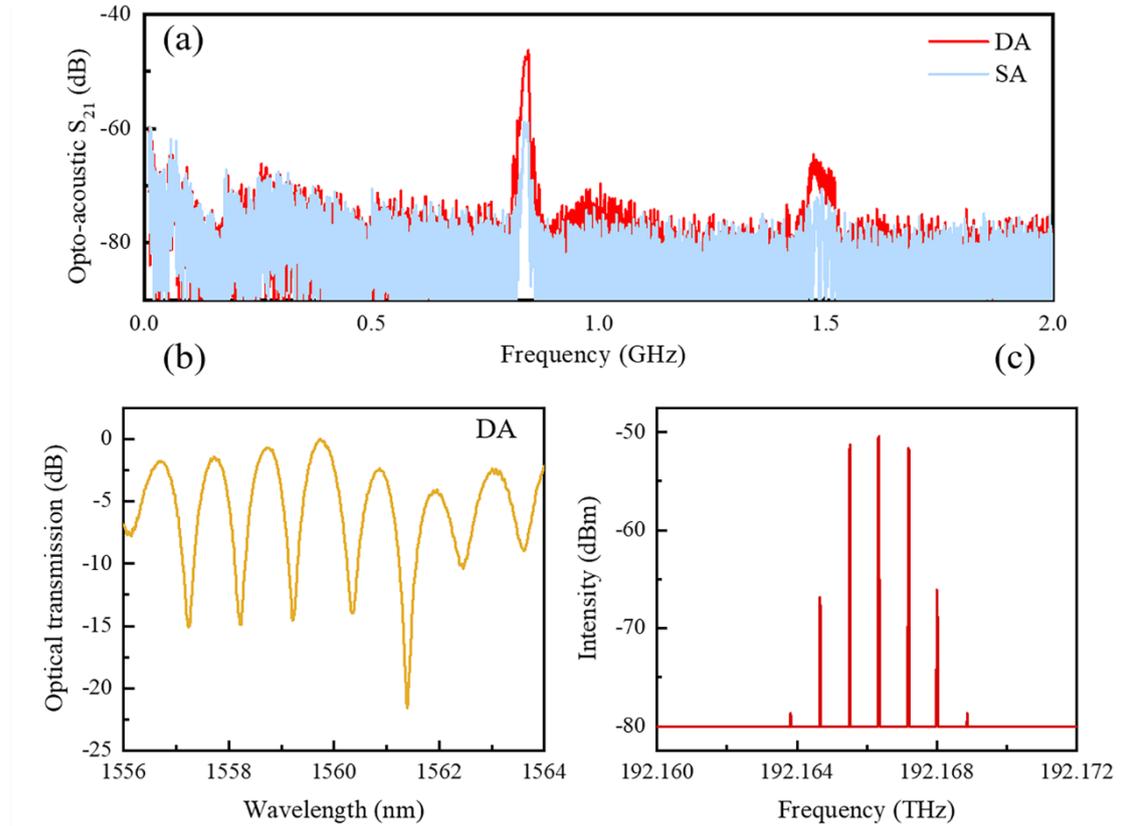

**Fig. 4 AO modulation. a** $S_{21}$ spectra of the TFLN-ChG hybrid MZI-based AO modulators with the SA and DA modulation configurations. **b** Normalized optical transmission spectrum of the AO modulator with the DA configuration. **c** Measured optical sidebands in the DA-modulated AO modulator at an RF power of 15 dBm.

**Demonstration of an OOK modulation link**

To demonstrate the low power consumption of the device, we construct an OOK modulation link using our nonsuspended built-in push-pull AO modulator. As shown in Fig. 5a, a microwave signal generator is used instead of the VNA to generate a square wave (OOK modulated signal) with a carrier wave frequency of 0.84 GHz and a modulation frequency of 1 MHz. The output light is amplified by an erbium-doped fiber amplifier, and then, the converted electrical signal is connected to a high-speed oscilloscope. Figure 5b shows the time-domain oscillogram of the modulated optical signal from the output port of our AO modulator using only an RF power of 6 dBm. The OOK modulated RF signal is loaded onto the DC optical carrier through our push-pull AO modulator, clearly demonstrating the microwave signal transmission capability of the developed on-chip AO modulator. Time-resolved oscillograms recorded for different RF powers are shown in Fig. 5c, and distortion of the sinusoidal modulated optical signal is gradually observed with increasing RF power due to the generation of high-order sidebands. Performing Fourier transformation on the time-domain signal under an RF power of 6 dBm in Fig. 5b, we can capture an up to third-order microwave beat-note signal, as shown in Fig. 5d, demonstrating the high efficiency merit of our developed built-in push-pull AO modulator.

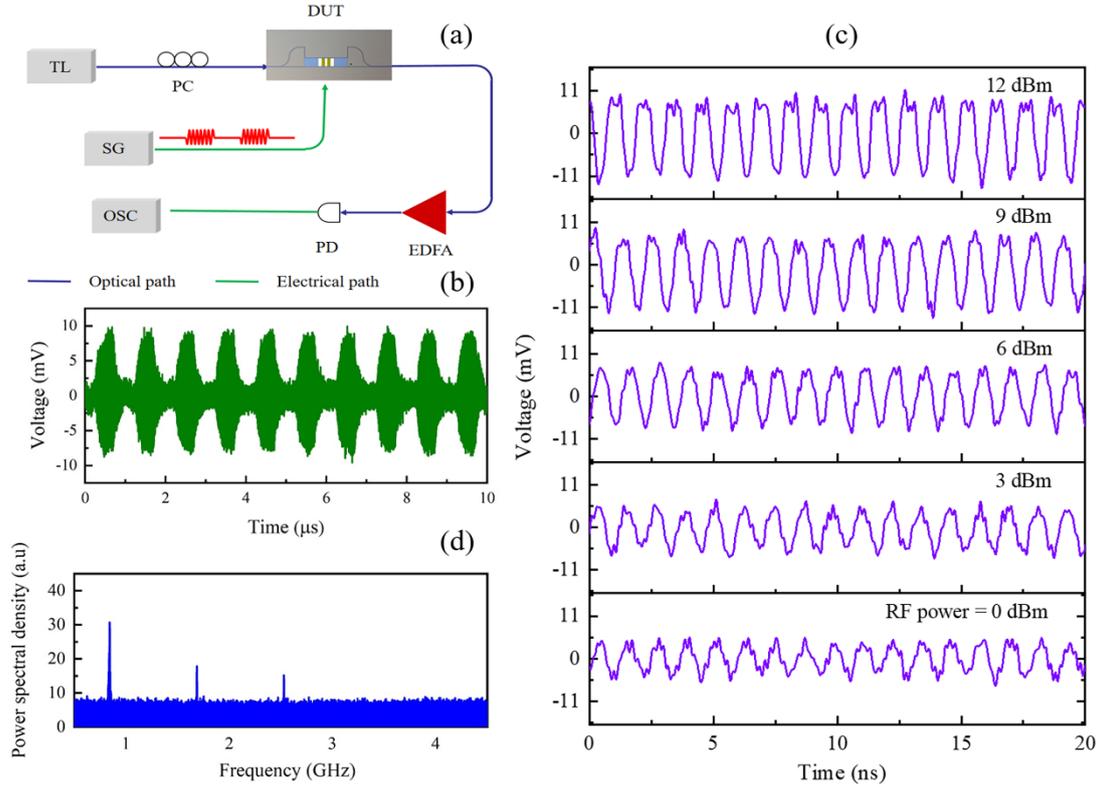

**Fig. 5 Experimental demonstration of an OOK modulation link using our built-in push-pull AO modulator. a** Schematic diagram of an OOK modulation link. The 1 MHz OOK modulated RF signal at 0.84 GHz is delivered from a signal generator to the device under test via a microwave probe. **b** Time-domain oscillogram of the amplitude-modulated optical wave sampled within 10 μs under 6 dBm RF power. **c** Measured time-resolved oscillograms recorded for different RF powers applied in our AO modulator. **d** Fourier transformation of the time-domain oscillogram corresponding to the input of 6 dBm RF power. SG: signal generator, EDAF: erbium-doped fiber amplifier, OSC: oscilloscope.

**Discussion**

Table 1 compares our SA-modulated and DA-modulated on-chip AO modulators based on nonsuspended TFLN-ChG hybrid waveguides with the present mainstream TFLN AO modulators. Good homogeneous TFLN AO modulators seriously depend on the construction of the suspended acoustic resonant cavites[22, 24], whereas our AO modulators avoid the fabrication of similar configurations, greatly relaxing the fabrication recipes of devices. Benefiting from the excellent acoustic wave response characteristics of the TFLN-ChG hybrid waveguide platform, the prominent $V_\pi L$ of 0.03 V cm is experimentally demonstrated based on our proposed built-in push-pull AO modulator. As a significant figure-of-merit (FOM), the $V_\pi L$ of our device with the DA configuration is one order of magnitude smaller than that of its counterpart based on a nonsuspended homogeneous TFLN waveguide platform without an acoustic cavity. This impressive modulation performance substantially originates from the increased overlap integral between the optical and acoustic waves. More specifically, the highly efficient bidirectional participation of the excited antisymmetric acoustic wave in the MZI two arms effectively enhances the energy proportion of the Rayleigh SAW at 0.84 GHz applied in the soft ChG rectangular waveguides. As another FOM, the phase shift per unit square root microwave power $\alpha_p$ is estimated to be 0.4 rad/√mW (see Supplementary note 5), which is desirable for high-performance AO modulators. In the future, we can further improve the microwave-to-optical conversion based on the nonsuspended TFLN-ChG hybrid waveguide platform by combining an optical resonator, an optomechanical resonant cavity, and a nanobeam waveguide with built-in push-pull IDTs.

Table 1. Comparison of modulation metrics for thin-film LN MZI-based AO modulators

| Ref./device | Platform | Acoustic cavity | Frequency (GHz) | $1-|S_{11}|^2$ (%) | L (μm) | $\alpha_p$ (rad/√mW) | $V_\pi L$ (V cm) |
|---|---|---|---|---|---|---|---|
| [23] | LN[a] | √ | 0.11 | 42 | 1200 | 0.073 | 2.5 |
| [31] | As$_2$S$_3$/SiO$_2$/LN[a] | √ | 0.11 | 35 | 2400 | 0.26 | 0.94 |
| [22] | LN[b] |  | 3.27 | 64 | 100 | 0.27 | 0.064 |
| [24] | LN[b] | √ | 1.16 | 19.3 | 45 | 0.54 | 0.019 |
| [24] | LN | √ | 1.9 | 50 | 45 | 0.017 | 0.38 |
| [24] | LN | x | 1.9 | 90 | 45 | 0.018 | 0.27 |
| Our work | SA: Ge$_{25}$Sb$_{10}$S$_{65}$/LN | x | 0.84 | 98.5 | 120 | 0.12 | 0.1 |
|  | DA: Ge$_{25}$Sb$_{10}$S$_{65}$/LN | x | 0.84 | 98.5 | 120 | 0.4 | 0.03 |

[a] In-plane metal grating reflectors were fabricated to construct an acoustic cavity. [b] Suspended TFLN was etched as an acoustic cavity. The MZI modulator configurations in Refs. [23, 31] are push-pull for Rayleigh SAWs, whereas for Refs. [22, 24], they are SA modulation for Lamb waves.

In conclusion, we propose and demonstrate a built-in push-pull AO modulator based on the nonsuspended TFLN-ChG hybrid waveguide platform. Benefiting from the excellent PE property of ChG and the antisymmetric acoustic mode excited by an impedance-matched IDT, the $V_\pi L$ of the push-pull AO modulator is measured to be as low as 0.03 V cm, reflecting the highly efficient modulation performance of our device with the DA configuration. Compared with the AO modulator with the SA modulation configuration, the DA-modulated counterpart has a twofold enhancement in the ME. In addition, the demonstration of SAW-driven AO modulation in the nonsuspended TFLN-ChG heterogeneous-integration waveguide platform extremely simplifies the fabrication recipe of the device and provides sufficient degrees of freedom to flexibly design high-performance on-chip AO interaction devices. As a verification of low power consumption, up to a third-order RF sideband was experimentally demonstrated via an efficient OOK modulation link. We anticipate that the development of a highly efficient on-chip AO modulator as a key component will offer opportunities for emerging RF-driven on-chip optical isolators and integrated analog optical computing devices.

**Materials and Methods**

## Device fabrication

The devices were fabricated on an X-cut thin-film LNOI wafer purchased from NANOLN, where the nominal thickness of the LN layer was 400 nm. We first deposited an 850 nm-thick $Ge_{25}Sb_{10}S_{65}$ membrane over the LNOI wafer by the thermal evaporation method. Then, we performed electron-beam lithography (EBL) to pattern the MZI waveguide structure as a mask using an electron-beam resist (ARP 6200.13) and transferred the photonic waveguide onto the $Ge_{25}Sb_{10}S_{65}$ film using reactive ion etching. Finally, the IDTs were fabricated through a lift-off process involving second-step EBL and gold deposition, where the thickness of the gold electrodes was 100 nm, with a 10 nm Ti adhesive layer previously deposited. The thickness of ARP6200.13 was controlled to 400 nm during spin coating using a spinning speed of 4000 rpm. A schematic of the fabrication processes of the device is shown in Supplementary note 6.

## Measurement methods

A C-band tunable laser (Keysight 8164B) was employed to measure the optical transmission spectrum of the MZI waveguide using a pair of lens fibers with a 3 μm mode field diameter. The characterization of the $S_{11}$ spectra for IDTs was conducted using a VNA (Keysight, N5225A) with the aid of a microwave probe (GGB, 40A-GSG-100-DP). Before the measurement of the $S_{11}$ spectrum was performed, the VNA was calibrated to reset the impedance of the cable and probe. By choosing a proper bias wavelength in the transmission spectrum of the MZI, the $S_{21}$ spectrum could be obtained with scanning of the microwave frequency in the VNA when the modulated optical wave was converted into an electrical signal via a high-speed photodiode (Newport, 1544-B). The spectrum of modulation sidebands was recorded by connecting the output fiber to a high-precision OSA (APEX, AP2088A). The acquisition of time-domain oscillograms corresponding to the modulated optical waves was completed by delivering amplified and filtered optical signals to an OSC (LeCroy, 80 GSa/s) after photoelectrical conversion through a photodiode.

## Acknowledgments

We acknowledge the funding support provided by the Key Project in Broadband Communication and New Network of the Ministry of Science and Technology (MOST) (No. 2019YFB1803904), the National Natural Science Foundation of China (Grant Nos. 62175095, 61805104, 62105377, U2001601, 61935013), the Science Foundation of Guangzhou City (202102020593), and the China Postdoctoral Science Foundation (2021M693599). L.W. acknowledges helpful discussions with Prof. Yuecheng Shen, Prof. Xiaojie Guo, and Dr. Shangsen Sun.

## Author contributions

L.W., Z.L., Z.Y., and W.Z. conceived the device design. S.Z. and D.L. carried out the device fabrication. Z.Y., W.Z. and M.W. performed the device measurements. L.W., Z.Y., W.Z., M.W., J.P. and H.L. carried out the data analysis. N.Z. provided the free-form design of the ultracompact 90° bending waveguide. All authors contributed to the writing. L.W. finalized the paper. L.W., W.L. and Z.L. supervised the project.

## Conflict of interest

The authors declare no competing financial interests.

**References**

1. Balram, K. C., Davanco, M. I., Song J. D. & Srinivasan, K. Coherent coupling between radio frequency, optical, and acoustic waves in piezo-optomechanical circuits. *Nat. Photonics*. **10,** 346-352 (2016).
2. Munk, D., et al. Surface acoustic wave photonic devices in silicon on insulator. *Nat. Commun*. **10,** 4214 (2019).
3. Jiang, W., et al. Efficient bidirectional piezo-optomechanical transduction between microwave and optical frequency. *Nat. Commun*. **11,** 1166 (2020).
4. Balram., K. C., Davanco, M. I., Song, J. D. & Srinivasan, K. Coherent coupling between radiofrequency, optical and acoustic waves in piezo-optomechanical circuits. *Nat. Photon*. **10,** 346-352 (2016).
5. Shao, L., et al. Non-reciprocal transmission of microwave acoustic waves in nonlinear parity-time symmetrical resonators. *Nat. Electronics*. **3,** 267-272 (2020) .
6. Sarabalis, C. J., et al. Acousto-optic modulation of a wavelength-scale waveguide. *Optica*. **8,** 477-483 (2021).
7. Li, H., Tadesse, S. A., Liu, Q. & Li, M. Nanophotonic cavity optomechanics with propagating acoustic waves at frequencies up to 12 GHz. *Optica.* **2,** 826-831 (2015).
8. Safavi-Naeini, A. H., Thourhout, D. V., Baets, R. & Laer, R. V. Controlling phonons and photons at the wavelength scale: integrated photonics meets integrated phononics. *Optica*. **6,** 213-232 (2019).
9. Schuetz, M. J. A., et al. Universal quantum transducers based on surface acoustic waves. *Phys. Rev. X*. **5,** 143-196 (2015).
10. Kittlaus, E. A., Otterstrom, N. T., Kharel, P., Gertler, S. & Rakich, P. T. Non-reciprocal interband Brillouin modulation. *Nat. Photonics.***12,** 613-619 (2018).
11. Kittlaus, E. A., et al. Electrically driven acousto-optics and broadband non-reciprocity in silicon photonics.*Nat. Photonics*. **15,** 43-52 (2021).
12. Tadesse, S. A. & Li, M. Sub-optical wavelength acoustic wave modulation of integrated photonic resonators at microwave frequencies. *Nat. Commun* **5,** 5402 (2014).
13. Fan, L., et al. Integrated optomechanical single-photon frequency shifter. *Nat. Photonics*. **10,** 766-770 (2016).
14. Dong, C. H., et al. Brillouin-scattering-induced transparency and non-reciprocal light storage. *Nat. Commun.* **6,** 6193 (2015).
15. Kim, J., Kuzyk, M. C., Han, K., Wang, H. & Bahl, G. Non-reciprocal Brillouin scattering induced transparency. *Nat. Phys*. **11,** 275-280 (2015).
16. Kowel, A. Acousto-optics-a review of fundamentals. *Proceedings of the IEEE*. **69,** 48-53 (1981).
17. Shao, L., et al. Integrated microwave acousto-optic frequency shifter on thin-film lithium niobate. *Opt. Express*. **28,** 23728-23738 (2020).
18. Tsai, C. S. Guided-wave acousto-optics: interactions, devices, and applications. (*Springer*, 1990).
19. Savage, N. Acousto-optic devices. *Nat. Photonics*. **4,** 728-729 (2010).
20. Fu, W., et al. Phononic integrated circuit and spin-orbit interaction of phonons. *Nat. Commun.* **10,** 2743 (2019).

21. Jr.-de Lima, M. M., Beck, M., Hey, R. & Santos, P. V. Compact Mach-Zehnder acousto-optic modulator. *Appl. Phys. Lett.* **89,** 121104 (2006).
22. Shao, L., et al. Microwave-to-optical conversion using lithium niobate thin-film acoustic resonators. *Optica.* **6,** 1498-1505 (2019).
23. Cai, L., et al. Acousto-optical modulation of thin film lithium niobate waveguide devices. *Photonics Res.* **7,** 1003-1013 (2019).
24. Hassanien, A. E., et al. Efficient and wideband acousto-optic modulation on thin-film lithium niobate for microwave-to-photonic conversion. *Photonics Res.* **9,** 1182-1190 (2021).
25. Qi, Y. & Li, Y. Integrated lithium niobate photonics. *Nanophotonics.* **9,** 1287-1320 (2020).
26. Boes, A., Corcoran, B., Chang. L., Bowers. J. & Mitchell, A. Status and potential of lithium niobate on insulator (LNOI) for photonic integrated circuits. *Laser Photonics Rev.* **12,** 1700256 (2018).
27. Ahmed, A. N. R., Shi, S., Zablocki, M., Yao, P. & Prather, D. W. Tunable hybrid silicon nitride and thin-film lithium niobate electro-optic microresonator. *Opt. Lett.* **44,** 618-621 (2019).
28. Rao, A., et al. High-performance and linear thin-film lithium niobate Mach-Zehnder modulators on silicon up to 50 GHz. *Opt. Lett.* **41,** 5700-5703 (2016).
29. Yu, Z. & Sun, X. Acousto-optic modulation of photonic bound state in the continuum. *Light. Sci. Appl.* **9,** 1-9 (2020).
30. Yu, Z. & Sun, X. Gigahertz Acousto-Optic Modulation and Frequency Shifting on Etchless Lithium Niobate Integrated Platform. *ACS Photonics.* **8,** 798-803 (2021) .
31. Khan, M. S. I., et al. Extraction of elastooptic coefficient of thin-Film arsenic trisulfide using a Mach–Zehnder acousto-optic modulator on lithium niobate. *J. Lightwave Technol.* **38,** 2053-2059 (2020).
32. Zhang, B., et al. On-chip chalcogenide microresonators with low-threshold parametric oscillation. *Photonics Res.***9,** 1272-1279 (2021) .
33. Lin, H., et al. Chalcogenide glass-on-graphene photonics. *Nat. Photonics.* **11,** 798-805 (2017).
34. Eggleton, B. J., Luther-Davies, B. & Richardson, K. Chalcogenide photonics. *Nat. Photonics.* **5,** 141-148 (2011).
35. Song, J., et al. Ultrasound measurement using on-chip optical micro-resonators and digital optical frequency comb. *J. Lightwave Technol.* **38,** 5293-5301 (2020).
36. Sohn, D. B., Orsel, O. E. & Bahl, G. Electrically driven optical isolation through phonon-mediated photonic Autler-Townes splitting. *Nat. Photonics.* **15,** 822-827 (2021).
37. Zhao, H., Li, B., Li, H. & Li, M. Scaling optical computing in synthetic frequency dimension using integrated cavity acousto-optics. arxiv.org/abs/2106.08494
38. Sun, S., et al. Inverse Design of Ultra- Compact Multimode Waveguide Bends Based on the Free- Form Curves. *Laser Photonics Rev*. **15,** 2100162 (2021).
39. Liu, Q., Li, H. & Li, M. Electromechanical Brillouin scattering in integrated optomechanical waveguides, *Optica.* **6,** 778-785 (2019).